\documentclass[11pt,a4paper]{article}
\usepackage{epsfig}
\RequirePackage{mathrsfs}

\newlength{\dinwidth}
\newlength{\dinmargin}
\def\eq#1{{Eq.~(\ref{#1})}}
\newcommand{\Le}{\left(}
\newcommand{\Ra}{\right)}

\newcommand{\beq}{\begin{equation}}
\newcommand{\eeq}{\end{equation}}
\newcommand{\beqar}[1]{\begin{eqnarray}\label{#1}}
\newcommand{\eeqar}{\end{eqnarray}}
\newcommand{\pd}[2]{\frac{\partial #1}{\partial #2}} 
 
\begin{document}

\title {{~}\\
{\Large \bf  Transport properties of a charged drop in an external electric field }\\}
\author{ 
{~}\\
{~}\\
{\large 
S.~Bondarenko$\,\,$\thanks{Email: sergeyb@ariel.ac.il},
K.~Komoshvili$\,\,$\thanks{Email: komosh@ariel.ac.il}} 
\\[10mm]
{\it\normalsize  Ariel University, Israel}\\}

\maketitle
\thispagestyle{empty}

\begin{abstract}
 Transport properties of a charged droplet of weakly interacting particles 
in an external field are investigated.
A  non-equilibrium distribution function which describes a process of the droplet transverse evolution with constant entropy in an external electric field is calculated. 
With the help of this distribution function, shear viscosity coefficients in the transverse plane are calculated as well. They are found to be very small and depend on the time of the droplet's expansion in a hydrodynamical regime and 
external field value.
An applicability of the results to the description of  initial states of quark-gluon plasma obtained in  high-energy interactions of nuclei is also discussed.

\end{abstract}

\newpage
\section{Introduction}
 Collisions of relativistic nuclei in the RHIC and LHC experiments at very high energies led to the revelation of a new state of matter named quark-gluon plasma (QGP). At the 
initial stages of the scattering, this plasma is assumed to be in a strongly interacting phase  and resembles as a liquid referred as a strongly coupled Quark Gluon Plasma (sQGP)
\cite{Shyr1}, whose microscopic structure is not well understood yet \cite{Shyr1,Shyr2,Berd,Koch,Liao,Nahrgang,Stein,Skokov,Rand}. Anyway, the data obtained at the RHIC experiments are in a good agreement with the predictions of the 
ideal relativistic fluid dynamics, \cite{Fluid,Fluid1}, which establish fluid dynamics as a main theoretical tool
to describe the collective flow in the collisions. As an input to the hydrodynamical evolution of the particles it
is assumed that after a very short time, $\tau\, <\, 1\,\, fm/s$ \cite{Fluid2},  the matter  reaches a thermal equilibrium and expands with a very small shear viscosity  \cite{Visc,Teaney}.

 In the process of the high-energy scattering, the thermal equilibrium may be achieved only for small fireballs 
 \cite{HotSp} of the matter \cite{Berd,Skokov,Teaney,HotSp1}; the whole colliding system cannot be in in a global equilibrium 
state because the nuclei
scattering at high energy is a highly non-equilibrium process \cite{Berd,Step}. Subsequent expansion of the matter's hot spot
occurs with the constant entropy \cite{Land1}, which justifies the applicability of the hydrodynamical description of the process. This adiabatic expansion continues till the value of the particle's mean free path becomes comparable with the
size of the system. In this stage, instead a liquid,  a gas of interacting particles whose density rapidly decreases reveals.

 Application of the fluid dynamics to the process of the fireball's expansion requires some initial conditions among which 
the most intriguing one is a small value of the shear viscosity/entropy ratio. Perturbative result for the shear viscosity calculations is large, \cite{PerVis}, and some new mechanisms of the explanation of the 
shear viscosity smallness are required. 
There are different approaches to the possible mechanisms of this smallness. Except the models of strongly interacting 
quarks and gluons \cite{Shyr1}, there are approaches based on the weakly interacting particles with novel mechanism
of the viscosity creation, see for example\cite{AnVis,AnVis1,AnVis2}. 
In this note we also propose some new mechanism responsible for the shear viscosity smallness, similar in some extent 
to the ideas of \cite{AnVis1,AnVis2}. Namely, 
we  consider a model of a small and very dense charged droplet, see \cite{Our1}, which inter-particle interactions are 
weak
similarly to the interactions of the asymptotically free quarks and gluons in QCD.
We investigate a hydrodynamical expansion of this unstable droplet and calculate a viscosity
of the process during the collisionless regime of the expansion.
Following to the \cite{AnVis1} we also call obtained viscosity as anomalous one, despite the fact that
the 
mechanism of the viscosity coefficients smallness in our approach is different from the proposed in \cite{AnVis1}.

 In our calculations we consider only an transverse expansion of small, \cite{Our1}, and dense droplet of charged particles. Following by \cite{Land1}, we consider the process 
of this expansion as the process with constant entropy.
Due to the fact that we consider a charged droplet, the distribution function of the system cannot be stationary, it must 
depend on time. Thereby, the expansion process of the charged droplet with the constant entropy is determined by  Vlasov's equation,
\cite{Klim}, which is 
 the main tool for the kinetic description of the process. We solve the Vlasov's equation  for the
time-dependent, non-equilibrium distribution function which being time-dependent anyway preserves constant value of the entropy. Moreover, we consider our hot spot in the transverse external field of the other, relativistic particles, see \cite{Our1}, which contributes to the Vlasov's equation solution. As we will see further, 
the interaction between the charged droplet and external field is the mechanism responsible for the small values of the  shear viscosity coefficients in the given framework.

  In the next Section 2, we derive the electromagnetic field potentials created by the relativistically moved charged 
drop. In Section 3, we write the Vlasov's equation for the charged drop in the external electric field,
whereas in Section 4, we rewrite the Vlasov's equation in a new, integral form. Also, in Section 4, we determine a new initial condition for the equation which is different from the given in Appendix A because of the presence of the external electric field. 
In further Section, Section 5, we calculate a non-equilibrium distribution function till the first order of the perturbative series, which is formulated and defined in this Section. In the Section 6 we investigate transport characteristic of the charged droplet in the external electric field, including transverse shear viscosity coefficients. The last Section 7 is a conclusion of the paper.

\section{An external field of relativistic charged drop}

To a first approximation, we consider a charged drop of the matter moving as whole with  some velocity. In the frame related with this drop, the distribution function of the droplet's particles is described by usual Vlasov's equations:
\beq\label{Vlas11}
\frac{\partial\,f_{s}}{\partial\,t}\,+\,\vec{v}\,\frac{\partial\,f_{s}}{\partial\vec{r}}\,+
\,\Le\,q\vec{E}\,+\,\frac{q}{c}\,\vec{v}\times\vec{B}\Ra\,\frac{\partial\,f_{s}}{\partial\vec{p}}\,=\,0\,.
\eeq 
with  Maxwell's equations
\beq\label{Vlas22}
\nabla\times\vec{E}\,=\,0\,,
\eeq
\beq\label{Vlas33}
\nabla\times\vec{B}\,=\,0\,,
\eeq
\beq\label{Vlas44}
\nabla\cdot\,\vec{E}\,=\,4\,\pi\,q\,n\,\int\,f_{s}(x,t)\,d^{3}v\,\,,
\eeq
\beq\label{Vlas55}
\nabla\cdot\,\vec{B}\,=\,0\,.
\eeq
In this approximation, for the usual Maxwellian initial distribution function,  the solution for the self-consistent field is the screened electrostatic potential
plus some constant  which can be considered further as an initial value of the potential:
\beq\label{ElStat}
\varphi\,=\,C_{0}\,\frac{e^{-r/r_{D}}}{r}\,+\,4\,\pi\,q\,n\,r_{D}^{2}
\eeq
see \cite{Our1,PScr} for example. Here $C_{0}\,$ is some constant and 
\beq
r_{D}^{2}\,=\,\frac{k_{B}\,T_{0}}{4\pi\,q^{2}\,n_{0}}\,\,
\eeq
is Debye length. 
Redefinig the potential's initial value we obtain finally the following expression for the potential
\beq\label{ElStat1}
\varphi\,=\,C_{0}\,\frac{e^{-r/r_{D}}}{r}\,
\eeq
which formally  can be regarded as originating from the exchange of "massive photons" with masses $M\,=\,1/r_{D}$.

 Now, let's consider the field of  \eq{ElStat1} in the rest frame of some another dense drop of  particles,
 which is in the rest relatively to the first drop. 
In this frame, the potential \eq{ElStat1} is created by fast moving dense cloud of the particles with the
volume $V_{0}$, particle's density $n$ and total charge $q\,n\,V_{0}\,$. In this reference
frame, therefore, the potential \eq{ElStat1} is described by  Proca equation  which has the 
following form:
\beq
\Delta\varphi\,-\,\frac{1}{c^2}\frac{\partial^2\,\varphi}{\partial^{2}\,t}\,-\,\frac{\varphi}{r^{2}_{D}}\,=\,
-\,4\pi\,q\,n_{0}\,V_{0}\,\delta(\vec{r}\,-\,\vec{r}_{s}(t))
\eeq  
Solution of this equation is well known:
\beq
\varphi(\vec{r},t)\,=\,\frac{q\,n_{0}\,V_{0}\,}{2\pi^{2}}\int\,d^{2}k_{\bot}\,e^{\imath k_{\bot}(r_{\bot}\,-\,b)}\,
\int_{-\infty}^{\infty}\,\frac{e^{\imath k_{z}(z\,-\,v\,t)}\,dk_{z}}{k_{\bot}^{2}\,+\,k_{z}^{2}\,\gamma\,+\,1/r_{D}^{2}}\,
\eeq
where $\vec{r}\,=\,(r_{\bot},z)\,$, the position of the moving drop is given by $\vec{r}_{s}\,=\,(b,vt)\,$ with
$r_{\bot}\,=\,(r_{x},r_{y})$\,, $b\,=\,(b_{x},b_{y})$ and $\gamma\,=\,\sqrt{1-\frac{v^2}{c^2}}$.
In the relativistic limit when $v\,\approx\,c$ we obtain in the first order  expansion over $\gamma$
for the electrostatic potential:
\beq
\varphi(\vec{r},t)\,=\,2\,q\,n_{0}\,V_{0}\,\delta(z\,-\,ct\,)\,K_{0}(|r_{\bot}\,-\,b|/r_{D})
\eeq 
where $K_{0}$ is Macdonald's function \cite{AbSteg}. Correspondingly,
for the vector potential of the electromagnetic field we have:
\beq
A_{z}(\vec{r},t)\,=\,\varphi(\vec{r},t)\,,
\eeq
\beq
A_{x}(\vec{r},t)\,=\,A_{y}(\vec{r},t)\,=\,0\,.
\eeq
After the gauge transformation with the gauge function
\beq
f\,=\,-\,2\,q\,n_{0}\,V_{0}\,\theta(z\,-\,ct)\,K_{0}(|r_{\bot}\,-\,b|/r_{D})
\eeq
we finally obtain the potentials of the moving drop in the rest frame of  another dense droplet:
\beq\label{Pot1}
\varphi(\vec{r},t)\,=\,A_{z}(\vec{r},t)\,=\,0\,,
\eeq
\beq\label{Pot2}
A_{x}(\vec{r},t)\,=\,-\,2\,q\,n_{0}\,V_{0}\,\theta(z\,-\,ct)\,\partial_{x}\,K_{0}(|r_{\bot}\,-\,b|/r_{D})\,,
\eeq
\beq\label{Pot3}
A_{y}(\vec{r},t)\,=\,-\,2\,q\,n_{0}\,V_{0}\,\theta(z\,-\,ct)\,\partial_{y}\,K_{0}(|r_{\bot}\,-\,b|/r_{D})\,.
\eeq
Here $\theta$ is the Heaviside step function and coordinate dependence of the potentials is factorized by two parts: the $\theta$ function depends only on the longitudinal $z$ coordinate whereas the second function depends on the transverse
$x$ and $y$ coordinates. There is a factorization of the dimensions. Therefore, further, we will consider only two dimensional dynamics of the droplet, taking   $v_{z}\,=\,0$ for the droplet in the rest.

\section{Vlasov's equation of the charged droplet in the external field}

  In this section, we consider an interaction between two different charged spots, one of which was created at the first stage of the interaction and stays in the rest and the second is moving towards the first one. 
Again, both droplets we consider as "rigid" drops, which interact at some initial moment of time. After the interaction, the second drop begins its expansion/compression in the transverse plain. We limit the dynamics in two dimensional plain, taking $v_{z}\,=\,0$ for the expanding/compressing droplet. 
This Vlasov's description of the droplet's dynamics is valid during some time $T$,
after which the collision processes between particles inside the droplet became to be important. The possible value of  $T$ we will estimate below.

  The system of equations \eq{Vlas11}, \eq{Vlas22} may be reformulated as a system of self-consistent field of the first spot in the field of the second. In this case instead \eq{Vlas11} we have:
\beq
\label{Vl1}
\frac{\partial\,f_{s}}{\partial\,t}\,+\,\vec{v}\,\frac{\partial\,f_{s}}{\partial\vec{r}}\,+
\,q\vec{E}_{total}\,\frac{\partial\,f_{s}}{\partial\vec{p}}\,=\,0\,.
\eeq 
Here we neglected magnetic field. The reason for that is simple.
The potentials \eq{Pot1}-\eq{Pot3} describe magnetic field with only non-zero azimuthal component:
\beq
B_{\theta}\,=\,\partial_{z}\,A_{r}\,,
\eeq
here $A_{r}(\vec{r},t)$
is a radial component of the potential, see \eq{Vl4} below.
The same is correct for the self-consistent magnetic field of the droplet.
Therefore, because in the 2-dimensional, transverse Vlasov's equation the main magnetic field contribution
in the force component has a form
$v_{z}\,B_{\theta}\,/\,c\,$,
when $v_{z}\,=\,0$, we obtain that only transverse electric field remains in the model.
In general it means, that in calculations we will neglect the radial, so called self-focusing, component of the
force in three dimensions. We argue in the conclusion that it does not affect on the main results of the manuscript.

 Thereby, in \eq{Vl1}  we have:
\beq
\vec{E}_{total}\,=\,\vec{E}_{s}\,-\,\vec{E}_{ext}\, 
\eeq
with  $\vec{E}_{s}\,$ as a new self-consistent field and $\vec{E}_{ext}$ as a  field of the second, incident drop.
Maxwell's equations for the self-consistent field in this case have a following form
\beq\label{Vl2}
\nabla\times\vec{E}_{s}\,=\,0\,,
\eeq
\beq\label{Vl3}
\nabla\cdot\,\vec{E}_{s}\,=\,4\,\pi\,q\,n\,\int\,f_{s}(x,t)\,d^{3}v\,.
\eeq
The vector potential which determines the $\vec{E}_{ext}\,$  field can be found by using 
\eq{Pot2}-\eq{Pot3} from the previous section. In cylindrical coordinates 
with the origin in the center of the rest spot, where we take  $\,r_{\bot}\,=\,r\,$,
we have:
\beq\label{Vl333}
\vec{E}_{ext}\,=\,-\,\frac{1}{c}\frac{\partial\,A_{r}}{\partial\,t}\,\hat{e}_{r}
\eeq
with the field
\beq\label{Vl4}
A_{r}(\vec{r},t)\,=\,\,A_{x}(\vec{r},t)\cos\,\theta + A_{y}(\vec{r},t)\,\sin\,\theta\,=\,-\,
2\, q\, n_{0}\,V_{0}\,\theta(z\,-\,c t)\, 
\pd{K_{0}(|r - b|/r_{0D})}{r}\,. 
\eeq
Defining an overall charge of the incident drop  equal to
\beq\label{Vl44}
Q_{T}\,=\,q\, n_{0}\,V_{0}\,,
\eeq
we have:
\beq\label{Vl444}
A_{r}(\vec{r},t)\,=\,-\,2\,Q_{T}\,\theta( \tau)  
 \pd{K_{0}(|r - b|/r_{0D})}{r}\,.
\eeq 
Here we denoted $\tau\,=\,ct\,$ because of the new rest frame with the origin in the droplet in rest,
therefore $\tau\,>\,0$ determines the evolution of the drop after the interaction with external field at $\tau\,=\,0$.

Also, denoting   $\xi_{0}\,=\,\frac{r}{r_{0D}}$, we obtain : 
\beq\label{Vl4444}
A_{r}(\vec{r},t)\,
 =-\,2\,\frac{Q_{T}}{r_{0D}}\, 
 \pd{K_{0}(\xi_{0})}{\xi_{0}}\,\theta( \tau)\,=\,
 F(r)\,\theta(\tau)\,.
\eeq
For the c.m.f. related with the droplet in the rest, we have $r\,=\,0$ and $\,b\,\rightarrow\,r$, so
introducing a cut-off around $r\,=\,0$ we denote in the following
$K_{0}(\xi_{0})\,=\,K_{0}(|r - r_{0D}|/r_{0D})$.
We underline, that the total charge of the incident drop $Q_{T}$ depends in general on a total energy of the process in the c.m.f. . 

 For the  Vlasov's equation \eq{Vl1}, redefining as well the time variable as $t\,\rightarrow\,ct\,=\,\tau$,
we obtain:
\beq
\label{Vl5}
\,c\,\frac{\partial\,f_{s}}{\partial\,\tau}\,+\,\vec{v}\,\frac{
\partial\,f_{s}}{\partial\vec{r}}\,+\,q\,\Le
\,\vec{E}_{s}\,-\,\vec{E}_{ext}\,\Ra\,
\frac{\partial\,f_{s}}{\partial\vec{p}}\,=\,0\,.
\eeq
with
\beq\label{Vl555}
\vec{E}_{ext}\,=\,\frac{\partial\,A_{r}}{\partial\,\tau}\,\hat{e}_{r}\,=\,F(r,s)\,\delta(\tau)\,\hat{e}_{r}\,,
\eeq
where $F(r,s)$ is the function from \eq{Vl444}.
In \eq{Vl5} there is no $z$ dependence neither in the distribution function nor in the electric field. The vectors $\vec{r},\vec{p}$
are radial there, therefore
\eq{Vl5} describes an evolution of the distribution function in two dimensional transverse plane 
as a function of the "time" parameter $\tau$. Also, we see, that because our distribution function does not depend on the 
azimuthal angle and because the external field is purely radial, the equation \eq{Vl5} in fact is one dimension with only radial 
dependence included.

\section{Integral form of Vlasov equation and new initial conditions}

In order to analyze an analytical solution of \eq{Vl5}-\eq{Vl555} it is more convenient to rewrite Vlasov's equation as 
an integral one. First of all, we rewrite this equation as following:\footnote{In our problem a velocity on the z axis is decoupled from the radial and angle velocities, i.e. $f_{s}\,\propto\,G(v_{z})\,f_{s}(v_{\theta}\,,\,v_{r})$ with $\int\,G(v_{z})\,d v_{z}\,=\,1\,$,}:
\beq
\label{Vl8}
\,\frac{1}{\zeta_{r}}\,\frac{\partial\,f_{s}}{\partial\,\tau}\,+\,\frac{
\partial\,f_{s}}{\partial r}\,=\,-\,\frac{q}{c \, \zeta_{r}}\,\Le
\,E_{rs}\,-\,E_{r ext}\,\Ra\,
\frac{\partial\,f_{s}}{\partial p_{r}}\,
\eeq
with $\zeta_{r}\,=\,v_{r} / c\,$\footnote{In the following we will denote $
\vec{\zeta}=(\vec{\zeta},\,\zeta_{\theta})\,=\,(\,v_{r} / c\,,\,v_{\theta} / c\,)$.}.
For the l.h.s. of this equation,  we find the fundamental solution from the following equation:
\beq
\,\frac{1}{\zeta_{r}}\,\frac{\partial\,f_{s}}{\partial\,\tau}\,+\,\frac{
\partial\,f_{s}}{\partial r}\,=\,\delta(\tau)\,\delta(r)\,,
\eeq
this solution is well known, see for example \cite{Vlad},
it is given by 
\beq\label{FSol}
\mathscr{E}\,=\,\,\zeta_{r}\,\theta(\tau)\,\delta(\,r\,-\,\zeta_{r}\,\tau\,)\,.
\eeq
Thereby, with the help of \eq{FSol}, we rewrite \eq{Vl8} as an integral equation:
\beq\label{Vl9}
f_{s}(r,\vec{\zeta},\tau)=-\frac{q}{c \, \zeta_{r}}\int dr^{'} d\tau^{'}
\mathscr{E}(\tau-\tau^{'},r-r^{'})\Le
E_{rs}(r^{'},\tau^{'})-E_{r ext}(r^{'},\tau^{'})\Ra
\frac{\partial\,f_{s}(r^{'},\vec{\zeta},\tau^{'})}{\partial p_{r}(\zeta_{r})}+
f_{0}(r-\zeta_{r}\tau ,\vec{\zeta})\,,
\eeq 
where the function $f_{0}(r,\vec{\zeta})$ will be determined later.
Inserting \eq{FSol} into \eq{Vl9}, we obtain\footnote{Here and in the following expressions firstly the derivative
over $p_{r}(\zeta_{r})$ is taken and after the value of $r-\zeta_{r}(\tau-\tau^{'})$ is inserting in
the expression for distribution function.} :
\begin{eqnarray}\label{Vl10}
&\,& f_{s}(r,\vec{\zeta},\tau) =
\,\nonumber\\
&\,& =
-\frac{q}{c}\int_{0}^{\tau}d\tau^{'}
\Le
E_{rs}(r-\zeta_{r}(\tau-\tau^{'}),\tau^{'})-E_{r ext}(r-\zeta_{r}(\tau-\tau^{'}),\tau^{'})\Ra
\frac{\partial\,f_{s}(r-\zeta_{r}(\tau-\tau^{'}),\vec{\zeta},\tau^{'})}{\partial p_{r}(\zeta_{r})}
+\,\nonumber\\
&\,& + f_{0}(r-\zeta_{r}\tau ,\vec{\zeta}).
\end{eqnarray}
With the help of expression for the self-consistent field 
\beq\label{Self}
E_{rs}(r,\tau)\,=\,\frac{4\,\pi\,q\,n\,}{r}\,\int^{r}\,d z\,z\,
\int\,f_{s}(z,v_{r},v_{\theta},\tau)\,d^{2}v\,,
\eeq
and for the external field
\beq\label{Ext}
E_{rext}(r,\,\tau)\,=\,-\,2\,\frac{Q_{T}}{r_{0D}}\,
\pd{K_{0}(\xi_{0})}{\xi_{0}}\,\delta(\tau)\,=\,2\,\frac{Q_{T}}{r_{0D}}\,
K_{1}(\xi_{0})\,\delta(\tau)\,,
\eeq
we obtain  for \eq{Vl10}:
\begin{eqnarray}\label{Vl11}
f_{s}(r,\vec{\zeta},\tau) & = & -\frac{4\,\pi\,q^2\,n\,}{r c}\int_{0}^{\tau}d\tau^{'}
\frac{\partial f_{s}(r-\vec{\zeta}(\tau-\tau^{'}),\vec{\zeta},\tau^{'})}{\partial p_{r}(\zeta_{r})}
\int^{r-\zeta_{r}(\tau-\tau^{'})} d z z
\int f_{s}(z,v^{'}_{r},v^{'}_{\theta},\tau^{'}) d^2 v^{'} +
\nonumber \\
& + &\frac{2\,q\,Q_{T} }{c\,r_{0D}}\,
K_{1}(\xi_{0}\,-\,\zeta_{r}\frac{\tau}{r_{0D}})\,
\frac{\partial\,f_{s0}(r-\zeta_{r}\tau,\vec{\zeta})}{\partial p_{r}(\zeta_{r})}
 + f_{0}(r-\zeta_{r}\tau ,\vec{\zeta}).
\end{eqnarray}
At $\tau\,=\,0$ this equation gives an equation for the $f_{s0}$:
\beq\label{InCon}
f_{s0}(r,p_{r})\,-\,\frac{2\,q\,Q_{T} }{c\,r_{0D}}\, 
K_{1}(\xi_{0}\,)\,
\frac{\partial\,f_{s0}(r,p_{r})}{\partial p_{r}(\zeta_{r})}\,=\,
f_{0}(r,p_{r})\,
\eeq
with some initial function
$f_{0}$. Therefore, \eq{Vl11} can be written in simpler form:
\begin{eqnarray}\label{Vl12}
f_{s}(r,\vec{\zeta},\tau)  & = &
\,-\,\frac{4\pi q^2 n}{r c} \int_{0}^{\tau}d\tau^{'}
\frac{\partial\,f_{s}(r-\zeta_{r}(\tau-\tau^{'}),\vec{\zeta},\tau^{'})}{\partial p_{r}(\zeta_{r})}
\cdot \nonumber \\
 & \, &
\cdot\int^{r-\zeta_{r}(\tau-\tau^{'})} d z z
\int f_{s}(z,v^{'}_{r},v_{\theta}^{'},\tau^{'}) d^2 v^{'}
+ f_{s0}(r-\zeta_{r}\tau ,\vec{\zeta})\,.
\end{eqnarray}
Equations \eq{InCon}-\eq{Vl12} are full analogues of the initial Vlasov equations with some initial condition defined by
\eq{InCon}.

An analytical solution of the \eq{InCon} can be easily found. For the case of non-relativistic
radial momenta, which is a case of interest and when $v_{r}\,/\,c\,<1\,$, it is
\beq\label{In22}
f_{s0}(r,\zeta_{r})\,=\,\frac{1}{\Lambda_{0}\,K_{1}(\xi_{0}\,)\,}\,e^{\zeta_{r}/(\Lambda_{0}\,K_{1}(\xi_{0}\,)\,)}\,\int_{\zeta_{r}}^{\infty}\,f_{0}(r,\zeta_{r}^{'})\,
e^{-\zeta_{r}^{'}/(\Lambda_{0}\,K_{1}(\xi_{0}\,)\,)}\,d\zeta_{r}^{'}\,,
\eeq
where
\beq\label{In18}
\Lambda_{0}\,=\,\frac{2\,q\,Q_{T}}{m c^2 r_{0D}}\,\,\approx\,
\frac{E^{ext}_{p}}{\mathscr{E}_{kin}\,}\,
\eeq
is the parameter which depends only on the field of the incident drop. Here $E^{ext}_{p}$ is a potential energy of the
interaction of the particle with the incident drop,
\beq
E^{ext}_{p}\,\propto\,\frac{\,q\,Q_{T}}{r_{0D}}\,,
\eeq
and $\mathscr{E}_{kin}$  is the relativistic kinetic energy of the incident drop.
Therefore, there are two different regimes in our problem, the high-energy (relativistic or weak external field limit) one when $\Lambda_{0}\,<\,1$
and another one of the strong external field when $\Lambda_{0}\,>\,1$. Solution \eq{In22} is valid in both cases but we are interesting in the 
high-energy regime when 
$\Lambda_{0}\,<\,1$ as $s\,\rightarrow\,\infty\,$\footnote{Here s is a squared total energy of the scattering process in the drop's rest frame.}. In this case,
we search our  function in the form of the following series:
\beq\label{In19}
f_{s0}(r,\vec{\zeta})\,=\,\sum_{i=0}^{\infty}F_{i}^{s0}\,\Lambda_{0}^{i}\,.
\eeq
From \eq{InCon} or from \eq{In22}, in the first two orders of the approximation, we obtain: 
\beq\label{In20}
f_{s0}(r,\vec{\zeta}) = f_{0}(r,\vec{\zeta})\,+\,\Lambda_{0}\,K_{1}(\xi_{0}\,)\,
\frac{\partial\,f_{0}(r,\vec{\zeta})}{\partial \zeta_{r}}\,.
\eeq
This form is useful for the case of high-energy (relativistic) limit of the problem for any
form of the initial distribution function $f_{0}(r,p_{r})$.

 Coming back to our initial differential formulation of Vlasov equation \eq{Vl5}, we see  that \eq{Vl12} 
is equivalent  to the usual Vlasov equation:
\beq
\label{Vl13}
\,c\,\frac{\partial\,f_{s}}{\partial\,\tau}\,+\,v_{r}\,\frac{
\partial\,f_{s}}{\partial r}\,+\,q\,
\,E_{rs}\,
\frac{\partial\,f_{s}}{\partial p_{r}}\,=\,0\,
\eeq 
with only self-consistent field included. The only difference of the new equation from the usual formulation is  the initial conditions, \eq{InCon}, where the influence
of the external field is included. Somehow it is very predictable result. Because of its factorized form, the external field may be considered as an additional source of the perturbation which acts only at initial time of the process of drop's compression/expansion by the external field.

\section{Solution of Vlasov's equation for "rigid-body" initial equilibrium distribution.}

\subsection{Initial condition and form of the perturbative expansion}

For the further 
calculations of the averaged values of the parameters of the problem.
 we need to solve Vlasov's equation and calculate distribution function. 
We consider the
integral equation \eq{Vl12}, obtained above, in the case
$\Lambda_{0}\,<\,1$. Solution of our initial Vlasov's equation, therefore, is given by
\eq{Vl5}. Considering this equation as  perturbative one, with "time" $\tau$ as a small parameter, we write:
\beq\label{PerExpan}
f_{s}(r,\,\vec{\zeta},\,\tau)\,=\,\sum_{i=0}^{\infty}\,f_{si}(r,\vec{\zeta})\,\tau^{i}\,.
\eeq
In the first order on $\tau$ we have
\beq\label{AppS1}
f_{s}(r,\vec{\zeta},\tau\,=\,0)\,=\,f_{s0}(r,\vec{\zeta})\,=\,f_{0}(r,\vec{\zeta})\,+\,\Lambda_{0}\,K_{1}(\xi_{0}\,)\,
\frac{\partial\,f_{0}(r,\vec{\zeta})}{\partial \zeta_{r}}\,,
\eeq
see \eq{In20}. The initial function $f_{0}(r,\vec{\zeta})$ we choose as a rotating "rigid-body" equilibrium distribution function described in  Appendix A. The reason for this choice is that this form of the distribution function is mostly appropriate for the configuration of the fields created in the high-energy scattering, see \cite{Davidson}.

\subsection{First order term of the distribution function}

In order to clarify a solution of our equation \eq{Vl12}, we perform Fourier transform over $r$ variable of our functions
of interest:
\beq
f_{s}(r,\,\vec{\zeta},\,\tau)\,=\,\int\,e^{\imath\,k\,r}f_{sk}(\vec{\zeta},\,\tau)\,dk\,.
\eeq
Taking into account 
the normalization of the initial distribution function, 
\beq
\int\,d^{2}\,v\,f_{s0}(r,\vec{v})\,=\,1\,,
\eeq
see \eq{In14}, the
equation \eq{Vl12}, therefore, acquires the following form:
\begin{eqnarray}
&\,&
\int  e^{\imath\,k\,r}\Le\,f_{sk0}(\vec{\zeta})\,+\,f_{sk1}(\vec{\zeta})\,\tau\,\Ra\,dk  = 
\nonumber \\
& = & -\,\Lambda\int_{0}^{\tau}\,\Le r-\zeta_{r}\tau^{'}\Ra^{2}\,d\tau^{'}
\frac{\partial}{\partial \zeta_{r}}\,\int\,e^{\imath\,k(r-\zeta_{r}\tau^{'})}\,f_{sk0}(\vec{\zeta})\,dk\, 
+\,\int\,e^{\imath\,k(r-\zeta_{r}\tau)}\, f_{sk0}(\vec{\zeta})\, dk\,,
\end{eqnarray}
with
\beq\label{LamDef}
\Lambda\,=\,\frac{2\pi q^2 n}{r m c^2}\,.
\eeq
Thereby, in the first order on $\tau$,
we obtain the following expression for the distribution function:
\beq
\tau\,\int\,e^{\imath\,k\,r} \,f_{sk1}(\vec{\zeta})\,dk  = \,-\,
\tau\,r^2 \,\Le\,
\frac{\partial}{\partial \zeta_{r}}\,\int\,e^{\imath\,kr}\,f_{sk0}(\vec{\zeta})\,dk\Ra\, 
-\,\tau\,\zeta_{r}\,\int\,e^{\imath\,kr}\,\Le \imath\,k\Ra \,f_{sk0}(\vec{\zeta})\, dk\,,
\eeq
that finally gives 
\beq\label{FOrSol}
f_{s1}(r,\,\vec{\zeta})\,=\,-\,\Lambda\,r^2\,
\frac{\partial\,f_{s0}(r,\,\vec{\zeta})}{\partial \zeta_{r}}\,
-\,
\zeta_{r}\,\frac{\partial\,f_{s0}(r,\,\vec{\zeta})}{\partial r}\,.
\eeq
Inserting expression \eq{AppS1} in \eq{FOrSol} , we obtain the first order distribution function term:
\begin{eqnarray}\label{FOrSol1}
f_{s1}(r,\,\vec{\zeta})& = &\,-\,\zeta_{r}\frac{\partial\,f_{0}(r,\,\vec{\zeta})}{\partial r}-
r^2\,\Lambda\,\frac{\partial\,f_{0}(r,\,\vec{\zeta})}{\partial \zeta_{r}}-
\,\zeta_{r}\,\Lambda_{0}\,
\frac{\partial\,K_{1}(\xi_{0})\,}{\partial r}
\frac{\partial\,f_{0}(r,\vec{\zeta})}{\partial \zeta_{r}}
\,\nonumber \\
\,& - &\,\zeta_{r}\,\Lambda_{0}\,K_{1}(\xi_{0})
\frac{\partial^{2}\,f_{0}(r,\vec{\zeta})}{\partial \zeta_{r}\partial r}-
r^2\,\Lambda\,\Lambda_{0}\,K_{1}(\xi_{0})\,\frac{\partial^{2}\,f_{0}(r,\vec{\zeta})}{\partial \zeta_{r}^2}\,.
\end{eqnarray}
The validity of our perturbative expansion is controlled by the request that
\beq\label{InEqu1}
r^2\,\Lambda\,\tau\,<\,1\,,
\eeq
all other parameters in the \eq{FOrSol1} are small. 
Now we can estimate the maximal value of $\tau$ in the problems as 
\beq
l_{0}\,=\,c\,T\,\propto\,\hbar\,/\,m c\,,
\eeq
see \cite{Land1}, and write $r^2\,\Lambda\,$ with the help of \eq{LamDef} as
\beq
r^2\,\Lambda\,\propto\,\frac{E_{pot}}{r m c^2}\,
\eeq
with $E_{pot}$ as a potential energy of the drop. 
Thereby, we can rewrite the inequality \eq{InEqu1} as
\beq\label{InEqu2}
E_{pot}\,<\,\frac{r}{l_{0}}\,m c^2\,.
\eeq
Thereby, we see from \eq{InEqu2}, that the inequality \eq{InEqu1} satisfies during 
all time of the applicability of Vlasov's equation, and, therefore, justify the expansion 
\eq{PerExpan} in whole "time" of interest.

\section{Transport properties of the drop}

In this section, we calculate the averaged velocities and shear viscosity of the drop with the help of
previously found distribution function. Namely, we will calculate the averaged radial velocity
\beq
\langle\,v_{r}\,\rangle\,=\,\frac{\int\,d^{2}v\,v_{r}\,f_{s}\,}{\,\int\,d^{2}v\,f_{s} }\,,
\eeq
the averaged azimuthal flow velocity
\beq
\langle\,v_{\theta}\,\rangle\,=\,\frac{\int\,d^{2}v\,v_{\theta}\,f_{s}\,}{\,\int\,d^{2}v\,f_{s} }\,,
\eeq
and the non-diagonal term of the stress-energy tensor required for the shear viscosity calculation
\beq
\sigma_{ij}\,=\,n\,m\,\int\,d^{2}v\,\Le\,v_{i}-\langle\,v_{i}\,\rangle\Ra\,\Le\,v_{j}-\langle\,v_{j}\,\rangle\Ra\,
\,f_{s}\,,
\eeq
where  $i,j\,=\,r,\theta$.

\subsection{Radial velocity calculations}

We calculate radial velocity up to the first order of perturbation series only, therefore
we take everywhere a density profile of the drop equal to
\beq
N\,=\,\int\,d^{2}v\,f_{s}=\int\,d^{2}v\Le f_{s0}+\tau\,f_{s1}\Ra=1-\Lambda_{0}
\frac{\partial\,K_{1}(\xi_{0})\,}{\partial r}
\int\,d^{2}v\,\zeta_{r}\,
\frac{\partial\,f_{0}(r,\vec{\zeta})}{\partial \zeta_{r}}=1+\tau\,\Lambda_{0}\,
\frac{\partial\,K_{1}(\xi_{0})\,}{\partial r}\,, 
\eeq
that gives for the first order term of the radial velocity 
\beq\label{RadV1}
\langle\,v_{r}\,\rangle\,=\,N^{-1}\,\int\,d^{2}v\,v_{r}\,f_{s}\,=\,N^{-1}\,\Le\,\int\,d^{2}v\,v_{r}\,f_{s0}\,+\,
\tau\,\int\,d^{2}v\,v_{r}\,f_{s1}\Ra\,=
\,N^{-1}\,\Le\langle\,v_{r}\,\rangle_{s}^{0}\,+\,\tau\,\langle\,v_{r}\,\rangle_{s}^{1}\Ra.
\eeq
Here we used the lower subscript "s" as the sign of averaging over the $f_{s}$ distribution function 
and the superscripts  denote the orders of the term in the perturbation series.

 The calculation of the first term in \eq{RadV1} is simple. We have:
\beq
\langle\,v_{r}\,\rangle_{s}^{0}\,=\,\int\,d^{2}v\,v_{r}\,f_{s0}\,=\,\int\,d^{2}v\,v_{r}\,\Le
\,f_{0}(r,\vec{\zeta})\,+\,\Lambda_{0}\,K_{1}(\xi_{0}\,)\,
\frac{\partial\,f_{0}(r,\vec{\zeta})}{\partial \zeta_{r}}\,\Ra\,.
\eeq
Opening the brackets we obtain:
\beq
\langle\,v_{r}\,\rangle_{s}^{0}\,=\,\langle\,v_{r}\,\rangle_{0}\,+\,c\,\Lambda_{0}\,K_{1}(\xi_{0}\,)\,
\int\,d^{2}v\,v_{r}\,\frac{\partial\,f_{0}(r,\vec{v})}{\partial v_{r}}\,.
\eeq
Taking into account the properties of our initial distribution $f_{0}$, we obtain finally:
\beq
\langle\,v_{r}\,\rangle_{s}^{0}\,=\,
-\,c\,\Lambda_{0}\,K_{1}(\xi_{0}\,)\,
\int\,d^{2}v\,f_{0}(r,\vec{v})\,=\,-\,\,c\,\Lambda_{0}\,K_{1}(\xi_{0}\,)\,.
\eeq 
We see, that as we assumed for the $\Lambda_{0}\,\langle\,\langle\,\,1$, our radial velocity is small $\langle\,v_{r}\,\rangle_{s}^{0}\,\langle\,\langle\,\,c$.
We obtained also, that the value of this velocity fully determined by the influence of the external 
field, and contrary to the results 
of Appendix A, the radial velocity is not zero even in zero order.

 In the first order of perturbation we have:
\beq\label{RadV2}
\langle\,v_{r}\,\rangle_{s}^{1}\,=\,\int\,d^{2}v\,v_{r}\,f_{s1}\,=\,-\,\int\,d^{2}v\,v_{r}\,\Le
\Lambda\,r^2\,
\frac{\partial\,f_{s0}}{\partial \zeta_{r}}\,+\,
\zeta_{r}\,\frac{\partial\,f_{s0}}{\partial r}\,
\Ra\,.
\eeq
Taking into account the full answer for the distribution function in this order \eq{FOrSol1}, we can rewrite 
\eq{RadV2} as
\beq\label{RadV3}
\langle\,v_{r}\,\rangle_{s}^{1}\,=\,\int\,d^{2}v\,v_{r}\,f_{s1}\,=\,-\,\int\,d^{2}v\,v_{r}\,\Le
\Lambda\,r^2\,
\frac{\partial\,f_{0}}{\partial \zeta_{r}}\,+\,
\zeta_{r}\,\frac{\partial\,f_{0}}{\partial r}\,
\Ra\,,
\eeq
all other terms gives zero contribution to the radial velocity value in this order.

The first term in the r.h.s. in \eq{RadV3} is
\beq
\Lambda\,r^2\,\int\,d^{2}v\,v_{r}\,\frac{\partial\,f_{0}}{\partial \zeta_{r}}\,=\,
c\,\Lambda\,r^2\,
\int\,d^{2}v\,v_{r}\,\frac{\partial\,f_{0}}{\partial v_{r}}\,=\,-\,
c\,\Lambda\,r^2\,\int\,d^{2}v\,f_{0}\,=\,-\,c\,\Lambda\,r^2\,,
\eeq
whereas the second term  is
\beq
\int\,d^{2}v\,v_{r}\,\zeta_{r}\,\frac{\partial\,f_{0}}{\partial r}\,=\,
\frac{1}{c}\,\frac{\partial\,}{\partial r}\,\int\,d^{2}v\,v_{r}^2\,f_{0}\,=\,
\frac{1}{c}\,\frac{\partial\,\langle\,v_{r}^{2}\,\rangle_{0}}{\partial r}\,.
\eeq
Here we need to calculate the square of the radial velocity averaged over \eq{In1} distribution function: 
\beq
\langle\,v^{2}_{r}\,\rangle_{0}\,=\,\int\,d^{2}\,v\,v_{r}^2\,f_{0}\,.
\eeq
We have:
\begin{eqnarray}
2\,\langle\,v^{2}_{r}\,\rangle_{0}\,
&=&
2\,\Le\frac{m}{2\,\pi}\Ra\,\int\,d^{2}\,v\,v_{r}^2\,
\delta\Le\frac{1}{2\,m}\Le\,p_{r}^{2}\,+\,(p_{\theta}\,-\,m\omega_{r} r)^{2}\,\Ra\,+\,\psi(r)\,
-\,k\,T_{\bot}\Ra\,\nonumber\\
& =&
\,\Le\frac{m}{2\,\pi}\Ra\,\int\,d^{2}\,v\Le v_{r}^2+v_{\theta}^{2}\Ra
\delta\Le\frac{1}{2\,m}\Le\,p_{r}^{2}\,+\,p_{\theta}^{2}\,\Ra\,+\,\psi(r)\,
-\,k\,T_{\bot}\Ra\,,
\end{eqnarray}
defining a new variable:
\beq
u^2\,=\,\frac{1}{2 m}\,\Le\, p_{r}^{2}\,+\,p_{\theta}^{2}\,\Ra\,,
\eeq
after the variable change we get the following expression:
\beq
2\,\langle\,v^{2}_{r}\,\rangle_{0}\,=\,\Le\frac{2}{m}\,\Ra\int\,d\,u^2\,u^2\,\delta(u^2\,+\,\psi(r)\,
-\,k\,T_{\bot})\,,
\eeq
which gives with the help of \eq{In111}:
\beq\label{RadVSq}
\langle\,v^{2}_{r}\,\rangle_{0}\,=\,\Le\frac{1}{m}\,\Ra\,\Le\,k\,T_{\bot}\,-\,\psi(r)\,\Ra\,=\,
\,\frac{k\,T_{\bot}}{m}\,\,\Le\,1\,-\,\frac{r^{2}}{r_{b}^{2}}\,\Ra\,.
\eeq
Taking both terms together, we obtain:
\beq\label{RadV7}
\langle\,v_{r}\,\rangle_{s}^{1}\,=\,c\,\Le
\frac{2\,k\,T_{\bot} r}{m c^2 r_{b}^{2}}\,+\,
\frac{2\pi q^2 n\, r}{ m c^2}\Ra\,=\,\frac{c\,r}{r_{b}^{2}}\,\Le
\frac{2\,k\,T_{\bot}}{m c^2 }\,+\,\frac{2\pi q^2 n\,r_{b}^{2} }{ m c^2}
\Ra\,.
\eeq

 We see, that we have here two possible different regimes. The first one
for the weakly interacting plasma, when
\beq
\Gamma\,=\,\frac{|\hat{U}_{p}|}{k T_{\bot}}\,<<\,1\,
\eeq
we obtain 
\beq\label{RadV4}
\langle\,v_{r}\,\rangle_{s}^{1}\,\approx\,c\,
\frac{2\,k\,T_{\bot} r}{m c^2 r_{b}^{2}}\,
\eeq
whereas for the strongly interacting plasma when
\beq
\Gamma\,>>\,1\,
\eeq
we obtain
\beq\label{RadV5}
\langle\,v_{r}\,\rangle_{s}^{1}\,\approx\,c\,\frac{2\pi q^2 n\, r}{ m c^2}\,.
\eeq
In both cases, the corrections are small due the large value of $m c^2$
in denominators of \eq{RadV4}-\eq{RadV5}. 

The final expression for radial velocity up to the first order, therefore, is the following:
\beq\label{RadV6}
\langle\,v_{r}\,\rangle\,=\,-\,N^{-1}\,\Le\,c\,\Lambda_{0}\,K_{1}(\xi_{0}\,)\,+\,\tau\,\frac{c\,r}{r_{b}^{2}}\,\Le
\frac{2\,k\,T_{\bot}}{m c^2 }\,+\,\frac{2\pi q^2 n\,r_{b}^{2} }{ m c^2}
\Ra\,\Ra.
\eeq
Taking 
\beq\label{NormF}
N^{-1}\,\approx\,1\,-\tau\,\Lambda_{0}\,
\frac{\partial\,K_{1}(\xi_{0})\,}{\partial r}\, 
\eeq
and keeping in final answer only liner on $\tau$ terms we obtain:
\beq\label{RadV22}
\langle\,v_{r}\,\rangle\,=\,-\,c\,\Lambda_{0}\,\Le\,\,K_{1}(\xi_{0}\,)\,-
\,\frac{\tau\,\Lambda_{0}\,}{r_{0D}}\,K_{1}(\xi_{0}\,)\,\frac{\partial\,K_{1}(\xi_{0})\,}{\partial \xi_{0}}\,\Ra
+\,\tau\,\frac{c\,r}{r_{b}^{2}}\,\Le
\frac{2\,k\,T_{\bot}}{m c^2 }\,+\,\frac{2\pi q^2 n\,r_{b}^{2} }{ m c^2}
\Ra\,.
\eeq
We see, that whereas the first term of expression \eq{RadV6} is fully determined by the external field and negative (compression process), the first order perturbation term has positive sign and describes the instability of the charged droplet
in absence of permanent external electric field.
Further, equation \eq{RadV2} we will use in the form similar to \eq{RadV1}:
\beq
\langle\,v_{r}\,\rangle\,=\langle\,v_{r}\,\rangle_{s}^{0}\,+\,\tau\,\langle\,v_{r}\,\rangle_{s}^{1}\,,
\eeq
where we group all liner on $\tau$ terms in the $\langle\,v_{r}\,\rangle_{s}^{1}\,$ term.

\subsection{Azimuthal velocity calculations}

We calculate the azimuthal velocity up to the first order
similarly to the radial velocity  calculations: 
\beq\label{AzV1}
\langle\,v_{\theta}\,\rangle\,=\,N^{-1}\,\int\,d^{2}v\,v_{\theta}\,f_{s}\,=\,N^{-1}\,\Le\,\int\,d^{2}v\,v_{\theta}\,f_{s0}\,+\,
\tau\,\int\,d^{2}v\,v_{\theta}\,f_{s1}\Ra\,=\,N^{-1}\,\Le\,\langle\,v_{\theta}\,\rangle_{s}^{0}\,+\,\tau\,\langle\,v_{\theta}\,\rangle_{s}^{1}\Ra\,.
\eeq
In the first order we have:
\beq\label{AzV11}
\,\langle\,v_{\theta}\,\rangle_{s}^{0}\,=\,\int\,d^{2}v\,v_{\theta}\,f_{s0}\,=\,
\int\,d^{2}v\,v_{\theta}\,\Le
\,f_{0}(r,\vec{\zeta})\,+\,\Lambda_{0}\,K_{1}(\xi_{0}\,)\,
\frac{\partial\,f_{0}(r,\vec{\zeta})}{\partial \zeta_{r}}\,
\Ra\,=\,\int\,d^{2}v\,v_{\theta}\,
\,f_{0}(r,\vec{\zeta})\,.
\eeq
We see that in this order, our answer is simply \eq{In7}:
\beq
\,\langle\,v_{\theta}\,\rangle_{s}^{0}\,=\,\langle\,v_{\theta}\,\rangle_{0}\,=\,\omega_{r}\,r\,.
\eeq
There is no influence of the external field in this order.

 In the next order, taking into account the first perturbated term, we have:
\beq\label{AzV2}
\,\langle\,v_{\theta}\,\rangle_{s}^{1}\,=\,\int\,d^{2}v\,v_{\theta}\,f_{s1}\,=\,-\,
\int\,d^{2}v\,v_{\theta}\,\zeta_{r}\,\Lambda_{0}\,K_{1}(\xi_{0})
\frac{\partial^{2}\,f_{0}(r,\vec{\zeta})}{\partial \zeta_{r}\partial r}\,.
\eeq
All other terms of \eq{FOrSol1} give zero contribution to the integral \eq{RadV2}.
Thereby we have:
\beq\label{AzV3}
\,\langle\,v_{\theta}\,\rangle_{s}^{1}\,=\,-\,\Lambda_{0}\,K_{1}\,\int\,d^{2}v\,v_{\theta}\,v_{r}\,
\frac{\partial^{2}\,f_{0}}{\partial v_{r}\partial r}\,=\,\Lambda_{0}\,K_{1}\,
\frac{\partial }{\partial r}\,\int\,d^{2}v\,v_{\theta}\,f_{0}\,=
\,\Lambda_{0}\,K_{1}\,\frac{\partial \langle\,v_{\theta}\,\rangle_{0}}{\partial r}\,.
\eeq
Taking both terms of \eq{RadV2}-\eq{RadV3} together and accounting \eq{NormF}  we obtain:
\beq\label{AzFin}
\langle\,v_{\theta}\,\rangle\,=\,\omega_{r}\,r\,\Le\,
1\,-\tau\,\Lambda_{0}\,
\frac{\partial\,K_{1}(\xi_{0})\,}{\partial r}\,\Ra\,
+\,\tau\,\Lambda_{0}\,K_{1}\,\omega_{r}\,.
\eeq
In this case the external perturbation leads to the increase of the droplet's rotation velocity.
We also rewrite this expression in more useful form. Collecting all liner on $\tau$ terms together
we rewrite \eq{AzFin} as
\beq
\langle\,v_{\theta}\,\rangle\,=\,\langle\,v_{\theta}\,\rangle_{s}^{0}\,+\,\tau\,\langle\,v_{\theta}\,\rangle_{s}^{1}\,.
\eeq 
This form of the answer we will use in the next subsection.

\subsection{Transverse shear viscosity of the drop }

 The shear viscosity is defined by the non-diagonal term of the stress-energy tensor. We have:
\beq
\sigma_{r\,\theta}\,=\,n\,m\,\int\,d^2\,v\,\Le v_{\theta}\,-\, \langle\,v_{\theta}\,\rangle \Ra\,
\Le v_{r}\,-\, \langle\,v_{r}\,\rangle \Ra\,f_{s}\,. 
\eeq 
In the zero order of our perturbation series, we have then
\beq
\sigma_{r\,\theta}^{0}\,=\,n\,m\,\int\,d^2\,v\,\Le v_{\theta}\,v_{r}\,-\,v_{r}\,\langle\,v_{\theta}\,\rangle_{s}^{0}\,
-\,v_{\theta}\,\langle\,v_{r}\,\rangle_{s}^{0}\,+\,\langle\,v_{r}\,\rangle_{s}^{0}\,\langle\,v_{\theta}\,\rangle_{s}^{0}\,\Ra\,f_{s0}\,=\,0\,,
\eeq
as it must be in the case of equilibrium.
In the  first order of perturbations we have:
\begin{eqnarray}
&\,&
\sigma_{r\,\theta}^{0}\,+\sigma_{r\,\theta}^{1}\,=\,
\,n\,m\,\int\,d^2\,v\,\Le v_{\theta}\,v_{r}\,-\,v_{r}(\langle\,v_{\theta}\,\rangle_{s}^{0}+\tau \langle\,v_{\theta}\,\rangle_{s}^{1})-
\right.
\\
& - &\,
\left.
v_{\theta}( \langle\,v_{r}\,\rangle_{s}^{0}+\tau \langle\,v_{r}\,\rangle_{s}^{1})\,+\,(\langle\,v_{r}\,\rangle_{s}^{0}+\tau \langle\,v_{r}\,\rangle_{s}^{1})(\langle\,v_{\theta}\,\rangle_{s}^{0}+\tau \langle\,v_{\theta}\,\rangle_{s}^{1})\,\Ra\,(f_{s0}+\tau f_{s1})\,.\nonumber
\end{eqnarray}
So we obtain:
\beq\label{Vis1}
\sigma_{r\,\theta}^{1}\,=\,n\,\tau\,\Le \langle\,v_{r} v_{\theta}\,\rangle_{s}^{1}\,-\,\langle\,v_{r}\,\rangle_{s}^{0}\,\langle\,v_{\theta}\,\rangle_{s}^{1}\,-\,
\,\langle\,v_{r}\,\rangle_{s}^{1}\,\langle\,v_{\theta}\,\rangle_{s}^{0}\,+\,
\langle\,v_{r}\,\rangle_{s}^{0}\,\langle\,v_{\theta}\,\rangle_{s}^{0}\,\int\,d^2\,v\,f_{s1}\,
\Ra\,.
\eeq
The only unknown term in the r.h.s. of the equation is the following one:
\beq
\langle\,v_{r} v_{\theta}\,\rangle_{s}^{1}\,=\,\int\,d^2\,v\,v_{\theta}\,v_{r}\,f_{s1}\,,
\eeq
with the $f_{s1}$ is given by  \eq{FOrSol1}. Keeping only non-vanishing after the integration terms we obtain:
\beq
\langle\,v_{r} v_{\theta}\,\rangle_{s}^{1}\,=\,-\,\int\,d^2\,v\,v_{\theta}\,v_{r}\,\Le
\,\zeta_{r}\frac{\partial\,f_{0}(r,\,\vec{\zeta})}{\partial r}+
r^2\,\Lambda\,\frac{\partial\,f_{0}(r,\,\vec{\zeta})}{\partial \zeta_{r}}
\Ra\,.
\eeq
Integration of this expression gives, for the first integrand:
\beq
-\,\int\,d^2\,v\,v_{\theta}\,v_{r}
\,\zeta_{r}\frac{\partial\,f_{0}(r,\,\vec{\zeta})}{\partial r}\,=\,-\,
\frac{\langle\,v_{\theta}\,\rangle_{s}^{0}}{c}\,\frac{\partial\,}{\partial r}\,\int\,d^2\,v\,v_{r}^{2}\,f_{0}\,=\,
-\frac{\langle\,v_{\theta}\,\rangle_{s}^{0}}{c}\,\frac{\partial\,\langle\,v_{r}^{2}\,\rangle_{0}}{\partial r}\,,
\eeq 
and for the second one:
\beq
-\,\int\,d^2\,v\,v_{\theta}\,v_{r}\,r^2\,\Lambda\,\frac{\partial\,f_{0}(r,\,\vec{\zeta})}{\partial \zeta_{r}}\,=\,
r^2\,\Lambda\,c\,\int\,d^2\,v\,v_{\theta}\,f_{0}\,=\,r^2\,\Lambda\,c\,\langle\,v_{\theta}\,\rangle_{s}^{0}\,.
\eeq
Taking into account \eq{RadV2}-\eq{RadV7}, \eq{AzV11}-\eq{AzV3} and summing up all terms we obtain :
\beq
\sigma_{r\,\theta}^{1}\,=\,\eta_{r\,\theta\,}^{\sigma\,\delta}\,
\frac{\partial \langle\,v_{\sigma}\,\rangle_{s}^{0}}{\partial r_{\delta}}\,=\,
\,n\,m\,\tau\,c\,
\Le\,\Lambda_{0}\,K_{1}\,\Ra^{2}\,\frac{\partial \langle\,v_{\theta}\,\rangle^{0}_{s}}{\partial r}\,+\,
3\,\,n\,m\,\tau\,\omega_{r}\,r\,\Lambda_{0}\,K_{1}\,
\frac{\partial \langle\,v_{r}\,\rangle_{s}^{0}}{\partial r}\,.
\eeq 
This expression  determines the shear viscosity coefficients in transverse plane:
\beq\label{Vis2}
\eta_{r \theta}^{ \theta r}\,=\,n\,m\,\tau\,c\,
\Le\,\Lambda_{0}\,K_{1}\,\Ra^{2}\,,\,\,\,\,
\eta_{r \theta}^{r r}\,=\,3\,n\,m\,\tau\,\omega_{r}\,r\,\Lambda_{0}\,K_{1}\,.
\eeq
This result is interesting. Both coefficients are proportional to the external field determined by 
the $\Lambda_{0}$ parameter which is small. Also, these coefficients are proportional to the $\tau$
and, therefore, stays  very small
during evolution of the system if $\tau$ remains small. Similarly to the definition in  \cite{AnVis1}
we can call this viscosity coefficients as anomalous.

 Indeed, the parameter  $\tau$  in \eq{Vis2} is not a mean free path as in  usual interpretations of the viscosity in 
different kinematic approximations.
The shear viscosity coefficients of \eq{Vis2} are determined by the collective expansion of the charged particles under an influence of the interactions of the particles inside the droplet and some initial perturbation caused by an external field.
Thereby, the expressions for this anomalous viscosity 
values are correct till
the Vlasov's approximation is correct in the description of the expanding dense matter. Therefore, the
 $\tau$ parameter in \eq{Vis2} is a  "time" of evolution of the droplet in
 the phase of the hydrodynamic expansion (compression)
 and the maximum value of this parameter is itself small. We estimate the value of 
$\tau$ approximately as:
\beq\label{Vis3}
\tau\,\propto\,\hbar\,/\,m\,c\,,
\eeq
see for example \cite{Land1}. Inserting \eq{Vis3} into \eq{Vis2} we obtain:
\beq\label{Vis4}
\eta_{r \theta}^{\theta r}\,\rightarrow\,\hbar\,n\,
\Le\,\Lambda_{0}\,K_{1}\,\Ra^{2}\,\approx\,
\hbar\,n\,\Le\frac{E^{ext}_{p}}{\mathscr{E}_{kin}}\Ra^{2}\,
\eeq
and
\beq\label{Vis44}
\eta_{r \theta}^{r r}\,\rightarrow\,2\,\hbar\,n\,\frac{\omega_{r} r}{c}
\frac{E^{ext}_{p}}{\mathscr{E}_{kin}}\,,
\eeq
where we took $K_{1}\,\propto\,1$ for all $r$ of interest.
We see, that the obtained coefficients change from  zero value till very small  maximum value
during the droplet's expansion. Indeed, the only large parameter in \eq{Vis4} is the density
of the droplet $n$, which can be very large, but $\Lambda_{0}$  is a small parameter of the approach and overall value of the viscosity is proportional to $\hbar$. Additionally, in the case when $\omega_{r}\,=\,0$ the second coefficient is
zero as well.

 Obviously, the viscosity/entropy ratio in our calculations remains small. Indeed, we consider Vlasov equation, the entropy during the process of the drop's expansion
remains constant and it equals to the initial entropy of the drop: $s\,=\,s_{0}\,=\,const$\footnote{Here s is an entropy of the process}.
Therefore, the ratio $\eta\,/\,s\,=\,\eta\,/\,s_{0}\,$
changes only because change of the viscosity coefficient $\eta$ and overall ratio changes from zero to some small value
determined by the viscosity coefficient value \eq{Vis4}-\eq{Vis44}.

\section{Conclusion}

 In this paper, we considered a process of the compression/expansion of the dense charged droplet in the
transverse plane under the influence of the external transverse electric field. The main motivation of our calculations was the investigation of the possibility of shear viscosity to entropy ratio smallness in the framework 
of some classical model. We considered a case of non-stationary transverse drop expansion/compression in the hydrodynamical phase with constant entropy. The main result of our calculations is given by \eq{Vis2}-\eq{Vis44} for the shear viscosity coefficients value, which really provides requested smallness.

The droplet we considered,  consists of charged particles, the external fields are not constant in  time, therefore the droplet cannot stay in the equilibrium state and begins to expand shortly after it's creation. The very early stage of the expansion of the droplet is  hydrodynamical one, \cite{Land1}, the collisions between particles are not important and the expansion happens with constant entropy. Thereby, the non-equilibrium, time-dependent distribution function of the droplet's evolution is determined by  Vlasov's equation with the external field included.

 We solve a classical problem for the distribution function of the  non-equilibrium system, therefore, first of all, it is instructive to check the self-consistency 
of the obtained solution. Beginning from the "rigid body" initial state, Appendix A, we can "guess" the influence of the external electric field on the droplet and compare it with the obtained answers. In the framework of the definitions of the paper, we  have that the case $\Lambda_{0}\,>\,0$ corresponds to the repulsive external force, that means droplet's external compression, and the case  $\Lambda_{0}\,<\,0$ corresponds to the attractive external force, that means droplet's external expansion, the case $\Lambda_{0}\,=\,0$ corresponds to the absence of the interaction.
In our case we consider  electromagnetic interactions, but in general, other types of interactions between the
droplets are possible in more complex models, see for example \cite{BFKL,Bond}.

  This processes of external compression/expansion are applied to the process of droplet's expansion under the influence
of charged particles inside the drop. From this point of view we see that our results are self-consistent and coincide with the naive expectations from the process. Indeed,
the radial velocity answer, \eq{RadV6}, consists of the two terms. The first one is due to the external field, depending on the sign of $\Lambda_{0}$ which can be negative, case of the compression, or positive, case of the expansion. The second term
in  \eq{RadV6} is positive and corresponds to the expansion of the droplet under the influence of the repulsive forces 
between charged particles in the drop. The azimuthal velocity expression, \eq{AzFin}, also consist of the two terms. The first one is the same as obtained for the initial distribution in Appendix A, while the second one is the correction to the initial value due to the external field. We see here, that for the case of the compression, the azimuthal velocity increases, and in the case of negative
$\Lambda_{0}$ it decreases, as it must be according with the classical point of view. The density profile of the drop
does not  remains constant as well. In the first order on $\tau$ it changes under an influence of the external field. The instability of the charged particles inside the drop influences only in the next orders on $\tau$.

It was already mentioned above, that the important result of the performed calculations is the expression for the shear viscosity coefficient \eq{Vis2}. 
It would be underlined, that the value of the coefficient is determined by the interaction of the 
droplet with the external field, i.e. due to the non-zero $\Lambda_{0}$ coefficient which depends on the strength of the interactions between  charged droplets\footnote{In the case when $\Lambda_{0}=0$ the viscosity should be determined by the
next order in the $\tau$ perturbative expansion of the distribution function and transport coefficients.}. Also, the "time" parameter
in \eq{Vis2} is the evolution "time" of the process, not the mean free path, and it's value is limited by the 
value of the "time's" applicability of the Vlasov's approximation, \eq{Vis3}, which itself is small. Thereby, our viscosity is time dependent and changing during the drop's expansion, but  it remains very small, also because
$\Lambda_{0}$ determined by \eq{In18} is a perturbative, small parameter of the initial conditions.
This mechanism of the viscosity coefficients smallness can be called as an anomalous one, similarly to 
\cite{AnVis1}. The entropy of the process remains constant during the process of the drop's expansion/compression.
It gives immediately that the ratio viscosity/entropy changes from zero to some value during the process of interest, but due to the smallness of the viscosity, this ratio anyway remains very small, see also \cite{BNL}.

 In our model we considered a  droplet, which we assume to be similar to the drops which are created in the very initial stage 
of the high-energy scattering of hadrons. Initially these drops are in the state of the thermal equilibrium and the smallness of the drop's size is required in this case, see \cite{Our1}. 
This required smallness of the drops leads to simple consequences which concern our calculations. In QCD such small drop will contain almost asymptotically free quarks and gluons, which interactions in this phase in some extend are similar to the electromagnetic interactions. Thereby, we can learn important lessons about QGP behavior even from the present, oversimplified calculations. Namely, if we assume that drops similar to considered are created in  high energy collisions, then the transverse
dynamics of the drops must be similar to the obtained here with only interaction coupling values different.
We see therefore, that the smallness of the shear viscosity to entropy ratio may be explained even in the framework of classical models for the systems of weekly interacting particles.

 In order to make the calculations clearer, we limited the consideration of the droplet's dynamic by the transverse plain only. In this case, the 
longitudinal dynamics is absent and, as was shown below,  magnetic fields may be neglected. 
Definitely, in this approximation, our results can be used in the case of three dimensional dynamics with
 some approximation only. 
However, the obtained form of Vlasov's equation \eq{Vl9} is
linear over the magnetic and electric fields and they are linearly independent in the expression. 
Therefore, the inclusion of magnetic field in the model  will definitely change the value of the radial force in  the final answers but we doubt that it will change also the properties and general structures of the obtained expressions
for the transverse plane. Some additional terms will be added to the answers after all. Our future calculations will include the 
magnetic field as well as the longitudinal dynamics, the present model we consider as a very preliminary step towards full dynamics description of the processes of interest.

In general we conclude thereby, that we performed the first step in the description of the evolution of  dense droplets
in the presence of external electromagnetic fields. 
This task is important because in some extend, our model can clarify a dynamics of interactions of 
quarks and gluons in the colored charged drops in QCD processes and mechanisms of shear viscosity smalness in it.
Also, we hope, that the microscopical details of
the hydrodynamical expansion of the charged drops may provide better understanding and
description of the data obtained in high-energy collisions of protons and nuclei in the LHC and RHIC experiment, \cite{Fluid},\cite{BNL}.

\newpage
\section*{Appendix A: A "rigid-rotor" initial
equilibrium distribution function}

\renewcommand{\theequation}{A.\arabic{equation}}
\setcounter{equation}{0}

Properties of initial distribution in our problem are defined by \eq{InCon}, which, in turn, depends
on some another function $f_{0}(r,\vec{p})$. As $f_{0}(r,\vec{p})$ we will choose a "rigid-rotor"
equilibrium function:
\beq\label{In1}
f_{0}(r,\vec{p})\,=\,\Le\frac{m}{2\,\pi}\Ra\,\delta(H_{\bot}-\omega_{r}\,P_{\theta}\,-\,k\,T_{\bot})\,G(p_{z})\,,
\eeq
where we have:
\beq\label{In2}
\int^{\infty}_{-\infty}\,G(p_{z})\,dp_{z}\,=\,1\,;
\eeq
\beq\label{In3}
\Le\frac{m}{2\,\pi}\,\int^{\infty}_{-\infty}\,dv_{\theta}\,\int^{\infty}_{-\infty}\,dv_{r}\,
\delta(H_{\bot}-\omega_{r}\,P_{\theta}\,-\,k\,T_{\bot})\Ra_{r\,=\,0}\,=\,1\,.
\eeq
the non-relativistic Hamiltonian of the problem is given by
\beq\label{In4}
H_{\bot}\,=\,\frac{1}{2\,m}\Le\,p_{r}^{2}\,+\,p_{\theta}^{2}\,\Ra\,+\,q\,\Phi_{s0}\,,
\eeq
where in \eq{In1} 
\beq\label{In5}
P_{\theta}\,=\,r\,\Le\,p_{\theta}\,-\,m\,\omega_{c}\,r\,/\,2\,\Ra\,,
\eeq
and
\beq\label{In6}
\omega_{c}\,=\,\frac{|q|\,B_{0}}{m\,c}\,
\eeq
is a cyclotron frequency in presence of some initial magnetic field in $z$ direction,
see \cite{Davidson}. An average azimuthal flow velocity
\beq\label{In7}
\langle\,v_{\theta}\,\rangle_{0}\,=\,\frac{\int\,d^{3}p\,v_{\theta}\,f_{0}\,}{\,\int\,d^{3}p\,f_{0} }\,=\,
\omega_{r}\,r\,
\eeq
is fully defined by some given angular velocity $\omega_{r}\,=\,const$\, that explains the name "rigid-rotor" 
equilibrium for the \eq{In1} function.

It is convenient to rewrite argument of delta function in \eq{In3} in the following form:
\beq\label{In8}
H_{\bot}-\omega_{r}\,P_{\theta}\,=\,
\frac{1}{2\,m}\Le\,p_{r}^{2}\,+\,(p_{\theta}\,-\,m\omega_{r} r)^{2}\,\Ra\,+\,\psi(r)\,,
\eeq
where an effective potential $\psi(r)$ is defined as
\beq\label{In9}
\psi(r)\,=\,\frac{m}{2}\Le \omega_{c}\,\omega_{r}\,-\,\omega_{r}^{2}\Ra\,r^{2}\,+\,q\,\Phi_{s0}\,.
\eeq
The corresponding Poisson equation for
the self-consistent potential $\Phi_{s0}$ can be solved in this case giving
\beq\label{In10}
\Phi_{s0}\,=\,-\,\pi\,n\,q\,r^2\,
\eeq
and we can rewrite potential \eq{In9} as
\beq\label{In11}
\psi(r)\,=\,\frac{m}{2}\Le \omega_{r}^{+}\,-\,\omega_{r}\,\Ra\,
\Le \omega_{r}\,-\,\omega_{r}^{-}\,\Ra\,r^{2}\,\,
\eeq
with
\beq\label{In12}
\omega_{r}^{\pm}\,=\,\frac{\omega_{c}}{2}\,\{1\,\pm\,\Le 1\,-\,n\,s_{q} \Ra^{1/2}\,\}
\eeq
and
\beq\label{In13}
s_{q}\,=\,\frac{8\,\pi\,q^{2}\,}{m\,\omega_{c}^{2}}
\eeq
Now we see that for the distribution function \eq{In1} the density profile is constant:
\beq\label{In14}
n(r) = \left\{
\begin{array}{rl}
& 1\,=\,const.,\,\,0\,\leq\,r\,<\,r_{b} \\
& 0 \,,\,\,\,\,\,\,\,\,\,\,\,\,\,\,\,\,\,\,\,\,\,\,\,\,\,\,\,\,\,  r\,\,>\,r_{b}.\\
\end{array} \right.
\eeq
with
\beq\label{In15}
r_{b}^{2}\,=\,\frac{2\,k\,T_{\bot}\,/\,m}{\Le \omega_{r}^{+}\,-\,\omega_{r}\,\Ra\,
\Le \omega_{r}\,-\,\omega_{r}^{-}\,\Ra}
\eeq
as some maximal radius of non-zero density. With the help of \eq{In15}, \eq{In11} acquires the following form:
\beq\label{In111}
\psi(r)\,=\,\frac{k\,T_{\bot}\,r^{2}}{r_{b}^{2}}\,.
\eeq
In the following we consider  the case when
\beq\label{In16}
r_{D}\,\leq\,r_{b}\,,
\eeq
that makes approximation \eq{In11} correct in the region of interest of our problem.


\end{document}